# Physics-integrated Neural Network for Quantum Transport Prediction of Field-effect Transistors

Xiuying Zhang[1,2], Linqiang Xu[3], Jing Lu[3], Zhaofu Zhang[4], Lei Shen[1*]

[1]Department of Mechanical Engineering, National University of Singapore, 9 Engineering Drive 1, Singapore 117575, Singapore.

[2]Hubei Key Laboratory of Electronic Manufacturing and Packaging Integration, Wuhan University, Wuhan 430072, China.

[3]State Key Laboratory for Artificial Microstructure and Mesoscopic Physics and School of Physics, Peking University, Beijing 100871, China.

[4]The Institute of Technological Sciences, Wuhan University, Wuhan 430072, China.

[*]Corresponding Author: shenlei@nus.edu.sg

**Abstract:**

Quantum-mechanics-based transport simulation is of well-known importance for the design of quantum-scale ultra-short channel field-effect transistors (FETs) with its capability of understanding the physical mechanism, while facing the primary challenge of the high computational intensity. Traditional machine learning is expected to accelerate the optimization of FET design, yet its application in this field is limited by the lack of both high-fidelity datasets and the integration of physical knowledge. Here, we introduced a physics-integrated neural network framework to accurately predict the transport $I_{ds}$-$V_{gs}$ curves of sub-5-nm gate-all-around (GAA) FETs using an in-house developed high-fidelity database. The transport curves in the database are collected from literature and our first-principles calculations. Beyond silicon, we include indium arsenide, indium phosphide, and selenium nanowires with different structural phases as the FET channel materials in our database. Based on the database, we built a physical-knowledge-integrated hyper vector neural network (PHVNN), in which five new physical features were added into the inputs for prediction transport characteristics of the devices, achieving a sufficiently low mean absolute error of 0.39. In particular, ~98% of the current prediction residuals are within one order of magnitude. Using PHVNN, we efficiently screened out the symmetric *p*-type GAA FETs that possess the same figures of merit with the *n*-type ones, which are crucial for the fabrication of homogeneous complementary metal-oxide-semiconductor (CMOS) circuits. Finally, our automatic differentiation analysis provides interpretable insights into the PHVNN, which highlights the important contributions of our new input parameters and improves the reliability of PHVNN. Our approach provides an effective method for rapidly screening appropriate GAA FETs with the prospect of accelerating the design process of next-generation electronic devices.

**Introduction**

Recently, ultrathin-body planer field-effect transistors (FETs), FinFETs, and gate-all-around (GAA) nanowire (NW) FETs have been proposed to improve the performance of ultra-short channel devicess[1]. Reduced thickness of both the channel and dielectric oxide in these nanometer FETs can decrease the effective screening length of the channel following $\lambda_{ch} \propto \sqrt{(\epsilon_{ch}/\epsilon_{ox})d_{ox}d_{ch}}$ and improve the gate electrostatics[2,3]. Here, $d_{ch}$ ($\epsilon_{ch}$) and $d_{ox}$ ($\epsilon_{ox}$) represents the thickness (permittivity) of the channel and dielectric oxide, respectively. Among them, the GAA NW-FET is considered having the best gate electrostatics due to the fully surrounding gate electrods[4] (**Fig. 1 (a)**). The lower-dimension channel of NW-FETs gives additional rise to the smaller $\epsilon_{ch}$[5] and leads to a smaller $\lambda_{ch}$. On top of that, the NWs have longer carrier mean free paths at room temperature[6] and higher carrier mobility than the corresponding two-dimensional (2D) or bulk forms[7], making GAA NW-FETs work with better performance than other state-of-the-art FETs. Besides, GAA NW-FETs can be monolithically stacked into a 3D device architecture, maximizing "packing density" and further increasing the transistor density[1].

Short-channel NW-FETs are expected to work close to the ballistic limit in theory[8]. The quantum effect must be considered in their transport properties[1,9]. Semiclassical modeling and simulation may result in large errors for the nanometer device due to the parameter dependence of transition matrix element[5]. Simulation models that focus solely on the channel segment of ultra-short channel transistors may fail to accurately capture their performance, as the electrode-channel contact plays a critical role[10]. First-principles-based quantum transport methods, such as nonequilibrium Green's function (NEGF), have widely proved its success in investigating the quantum transport properties of nanometer devices. However, such high-fidelity simulation approaches are usually computationally expensive and time-consuming. A fast and accurate model for the simulation and design optimization of GAA NW-FET is yet to be developed.

To address these challenges, several works on the basis of neural network (NN) have been proposed to simplify the ultra-short device simulation process[11-18], including physics-based NNs to improve the accuracy[18-22]. Nevertheless, these NN-frameworks are trained on datasets from classical or semiclassical simulation[23,24], and limited to the silicon (Si) FETs with channel length larger than 10 nm. The requirement of massive amounts of training data hinders the application of NN models in ultra-short channel GAA NW-FETs, provided the quite limited quantum transport data available. NNs with prior physical knowledge as inputs are expected to address the limited-dataset issues[19]. Furthermore, regarding the interpretation of the black-box NNs, the challenge lies in identifying which descriptors are important for transport character prediction, and how they are related to the properties involved in the modelling. NNs with prior physical knowledges would highly reduce the data acquisition time and the required input parameters. In such circumstances, the descriptor importance analysis becomes easier and the model more explainable.

In this work, we constructed a quantum transport simulation database, which contains over one hundred high-fidelity transport characters for not only Si, but also indium arsenide (InAs), indium phosphide (InP), and selenium (Se) sub-5 nm NW-FETs (**Fig. 1 (b-f)**). The database is gained from both literatures and our first-principles NEGF calculations, where the ballistic transport is considered. Following this, we developed a physical-knowledge-integrated hyper vector NN (PHVNN) for predicting the transport characteristics of the FETs. The output current ($I_{ds}$) sequences from the transport curves, which are 16-dimentional hyper vectors, are chosen as the output. Several prior physical knowledge like forbidden gap ($E_g$) and effective mass ($m^*$), are incorporated into the device descriptors as inputs in the PHVNN, as they significantly influence the carrier transport. In the presence of these physical parameters, the prediction error of log($I_{ds}$) vectors is significantly reduced from 0.65 to 0.39, making about 98% of the $I_{ds}$ residuals in the range of one order of magnitude. With the help of PHVNN, we efficiently obtained a few pair of *n*- and *p*-type FETs that have the same figures of merit, holding the promise of further development of *p-n* junctions. Physical parameter importance analysis and automatic differentiation analysis are finally conducted. The consistent analysis results with prior physical principles demonstrate the accuracy and reliability of our model. The PHVNN is capable of greatly reducing the simulation time of a transport curve for ultra-short channel GAA NW-FETs. Thereby, the model will significantly improve development efficiency and shorten the development cycle of GAA NW-FETs, while providing robust data support for further device design.

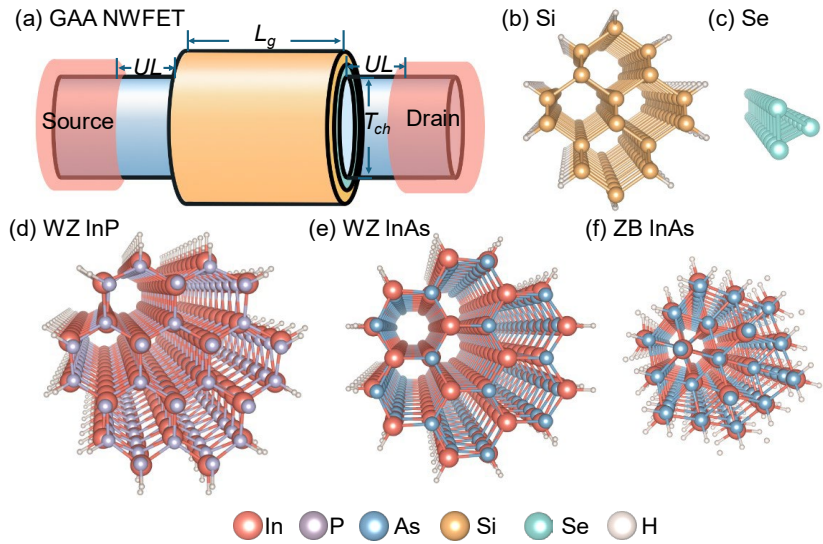

**Fig. 1**. (a) Schematic diagram of a GAA NW FET. Atomic structure of Si (b), Se (c), WZ InP (d), WZ InAs (e), and ZB InAs (f) NWs. Among them, the Si, InP, InAs NWs have hydrogen passivation.

## Results
### 1. Development of Sub-5 nm GAA NW-FET Database

The lack of dataset limits the application of machine learning (ML) in the design of short-channel GAA NW-FETs. Here, we first present a quantum transport database for sub-5 nm GAA FETs, mainly including the $I_{ds}$-$V_{gs}$ curves. These $I_{ds}$-$V_{gs}$ curves originate from two sources: literatures (majority) and our own calculations. To maintain consistency, all of them were generated using the first-principles NEGF method within the Quantum Atomistic ToolKit (Quantum ATK) package[25]. Due to the ultra-short channel lengths, an ideal ballistic electronic transport approach is employed for device simulation. The details of the calculations are consistent with the corresponding material studies[5,26-28], ensuring consistent settings when searching for symmetric FETs in homogeneous complementary metal-oxide-semiconductor (CMOS) design.

The subthreshold swing (*SS*) and transconductance ($g_m$) are two key quantitative descriptors used to assess gate controllability. Both *SS* and $g_m$ are extracted from the $I_{ds}$-$V_{gs}$ curves (**Fig. 2 (a, b)**). Therefore, exploring the $I_{ds}$-$V_{gs}$ curves is essential for evaluating the transport performance of devices and guiding their design. In our database, the maximum $I_{ds}$ is $4.8×10^3$ μA/μm and $6.7×10^3$ μA/μm for *n*-type and *p*-type transistors, respectively. The minimum $I_{ds}$ is $7.4×10^{-6}$ μA/μm and $1.1×10^{-5}$ μA/μm for *n*-type transistors and *p*-type transistors, respectively. The SS values in the database range from 64 mV/dec to 275 mV/dec approximately[26,27]. Such a rich and high-fidelity database provides a broad foundation for the exploration of GAA NW-FET, enabling extensive performance exploration and device design for NW-FETs.

*Materials parameters* - Our database is not limited the central materials to Si nanowire but also III-V nanowires like InAs[28-30] and InP[26,31,32], as well as Se NWs[27] (**Fig. 1 (b-f)**) with different nanowire diameters. III−V materials occur in either the zincblende (ZB) or wurtzite (WZ) crystal structure[33,34], and both of these two phases of InAs are included in our database. Specifically, the database comprises 109 transistors, including 25 Si NW-FETs, 17 WZ InAs NW-FETs, 16 ZB InAs NW-FETs, 27 InP NW-FETs, and 24 Se NW-FETs. This variability of channel materials provides the database with abundant of material characters. The channel diameter ($d_{ch}$) varies significantly, from as small as approximately 0.4 nm for Se NWs to as large as around 1.6 nm for InP NWs (**Fig. 2 (c)**). This variation in materials leads to a broad spectrum of materials properties in the database, such as the carrier effective mass ($m^*$) and band gaps ($E_g$) (**Fig. 2 (d, e)**). The effective mass varies dramatically, from as low as ~0.1 $m_0$ (Si CBM)[5] to as high as ~4.4 $m_0$ (WZ-InP CBM), while the band gaps span from a narrow ~1.6 eV (WZ-InAs)[28] to a wide ~3.4 eV (WZ-InP).

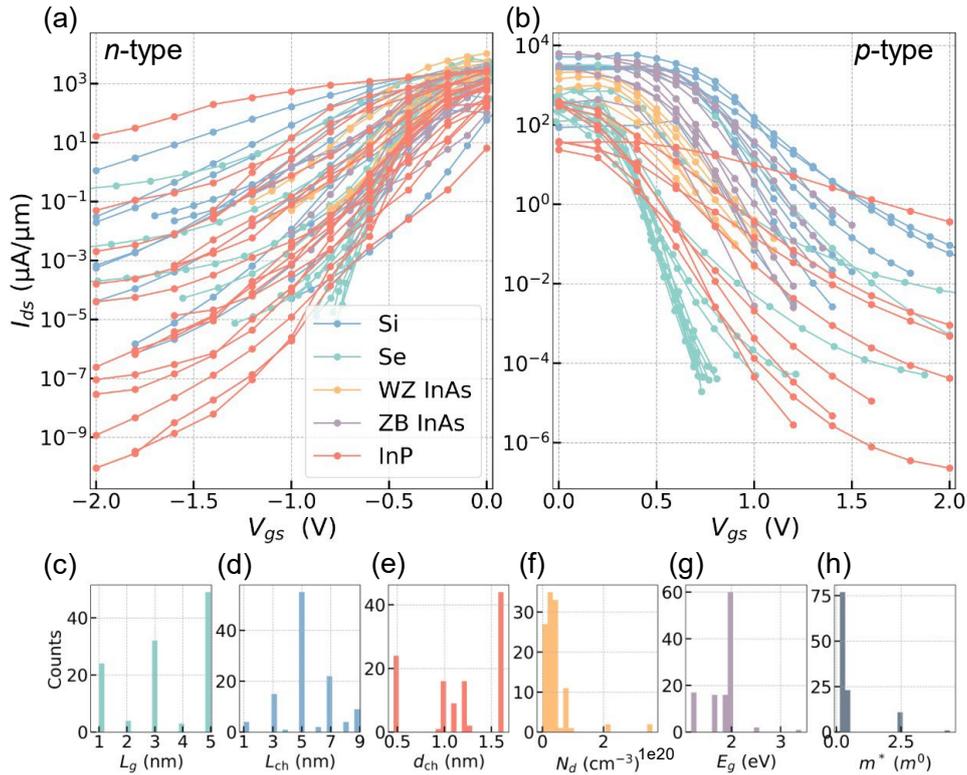

**Fig. 2**. $I_{ds}$-$V_{gs}$ curves of the *n*- and *p*-type GAA NW-FETs (a, b) and data distribution of the $d_{ch}$ (c), $E_g$ (d), $m^*$ (e), $L_g$ (f), $L_{ch}$ (g), and $N_d$ (h) in the database.

*Device parameters* - The FET device contains the source/drain electrodes and a central region. The electrodes are doped with electrons or holes at varying concentrations ($N_d$) to create *n*-type or *p*-type FETs, with our database including 64 *n*-type NW-FETs and 45 *p*-type NW-FETs. We specifically calculated the transport curves of p-type InP NW-FETs with a doping concentration of $3.6\times10^{19}$ cm$^{-3}$ ourselves to ensure data balance between n- and p-type. The central region consists of the gate region and two underlap (UL) regions (**Fig. 1 (a)**). The UL regions are adopted to enhance the device performance by decreasing the impact of the electrodes on the channel and reducing the leakage current[5]. The gate region is firstly covered with dielectric oxide, usually SiO$_2$ or HfO$_2$, followed by a metal layer to control the current. The channel length ($L_{ch}$) is thus the sum of the gate length ($L_g$) and two UL length ($L_{UL}$). In our NW-FET design, the gate lengths range from 1 nm to 5 nm (**Fig. 2 (f)**), while the UL lengths range from 0 nm to 5 nm, resulting in an overall channel length range of 1 nm to 9 nm (**Fig. 2 (g)**). The electrode doping concentration ($N_d$) differs significantly across the devices (**Fig. 2 (h)**). This variation in doping concentration causes shifts in the Fermi level, affecting the number of carriers that can transport across the device. Such differences in FET design parameters allows for a more comprehensive exploration of device performance across different material systems. The operating principle of a typical MOSFET is its gate controllability.

## 2. Physical-knowledge-integrated Hyper Vector Neural Network

To address the complexities of the $I_{ds}$-$V_{gs}$ curves with fewer data restriction, we develop a hyper vector NN model integrated with physics-knowledge. It is worth noting that directly predicting the $I_{ds}$ for each $V_{gs}$ individually tends to overfit because the data points in an $I_{ds}$-$V_{gs}$ curve are strongly correlated. Instead, the predicting the entire $I_{ds}$ sequence for the whole $I_{ds}$-$V_{gs}$ curve is more reasonable. The presence of low but strictly positive $I_{ds}$ values in the database imposes additional constraints on the model's parameter space. To address these two challenges, our model uses the logarithmic values, log($I_{ds}$) sequences, as output, which aligns with common practices in semiconductor research. The log($I_{ds}$) sequences are high-dimensional or hyper vectors. A grid search is then performed for hyperparameter optimization, varying the number of hidden layers, the number of neural nodes in the hidden layer, and the active function. Our testing shows that the model with one hidden layer with neural nodes of 32 and active function of tanh has the best performance, achieving the lowest test set mean absolute error (MAE) while avoiding overfitting.

The rational selection of input parameters is crucial for ensuring prediction accuracy, especially when working with fewer training datasets. Typically, the device size parameters, such as gate length, channel diameter, and equivalent oxide thickness (EOT) along with transport settings such as bias voltage, and doping concentration, are used as input features for the NNs[20]. However, parameters related to the channel materials, including the bandgap ($E_g$), effective mass ($m^*$) and the carrier Fermi velocity ($v_F$), also play an important role in determining the device performance[26,28,35]. Additionally, electrode properties, which will be affected by doping, cannot be fully captured by the doping concentration alone. Therefore, we also incorporate doping-induced changes in electrode parameters, including the Fermi level shifting ($\mu_F$) and contributed carriers to the transition per atom ($N(E)$) (**Fig. S1**), as input features in this work (**Table 1**). The details regarding the input parameters and the NN architecture can be found in the Methods section and Supporting Information.

To evaluate the model performance with the five new physical parameters, we conducted an ablation study[36] as shown in **Fig. S2** with the results summarized in **Fig. 3(a)**. The model is trained based on our transport curve database. As some of the transport curves for Se NW-FETs are significantly overlapped, we removed some of them to reduce the risk of model overfitting. Ultimately, 100 $I_{ds}$-$V_{gs}$ curves were selected for model training and were split into training and test sets with a 9:1 ratio. In all ablation tests, the device size parameters and the transistor setting parameters were included as group *X* in **Table 1** and **Fig. 3(a)**, and only the five physical parameters are varied. Here, the five physical parameter, $E_g$, $m^*$, $v_F$, $\mu_F$, and $N(E)$ (**Table 1**), are represented as *a*, *b*, *c*, *d*, and *e* in **Fig. 3 (a)**, respectively. The MAE for the test set was 0.65 when none of the physical parameters were included. The significant difference in MAE between the training and validation sets indicates overfitting (**Fig. S2**). After incorporating all five physical parameters, the test MAE is decreased to 0.39, and the difference between the training and validation sets became almost indistinguishable. The models with equal or smaller test MAEs are discussed in Physical analysis section. The improved performance of the NN underscores the importance of these carefully selected physical parameters in capturing carrier transitions in FETs. Ultimately, our ANNs utilizes an 11-dimensional vector (**Table 1**) as input for the $I_{ds}$ hyper vectors of GAA NW-FETs prediction.

**Table 1**. The input parameters used in the PHVNN.

|  | Parameters |  |
| --- | --- | --- |
| Group *X* | Gate Length | $L_g$ |
|  | Under lap | $L_{UL}$ |
|  | Channel diameter | $d_{ch}$ |
|  | EOT thickness | $d_\epsilon$ |
|  | Bias voltage | $V_{ds}$ |
|  | Doping concentration | $N_d$ |
| Five new proposed physical parameters | Bandgap | $E_g$ |
|  | Effective mass | $m^*$ |
|  | Carrier Fermi velocity | $v_F$ |
|  | Fermi level shifting | $\mu_F$ |
|  | Carrier contribution states | $N(E)$ |

The residuals of $\log(I_{ds})$ between the predicted and quantum calculated values for the transistors in the database was also calculated to assess the prediction accuracy (**Fig. 3 (b)**). About 98% of the $\log(I_{ds})$ residuals falls within the range of [-1.0, 1.0], indicating that the prediction error for $I_{ds}$ fluctuates within an order of magnitude. This level of precision is quite accurate for many practical purposes in semiconductor research. The $I_{ds}$ residuals distribution within the range of [-2.5, 2.5] (**Fig. 3 (b)**, inset), which aligns with the test MAE of the PHVNN ($10^{0.39} \sim 2.45$), also shows a high concentration around zero. The high concentration of $I_{ds}$ residuals indicates that the model consistently produces results very close to those obtained from quantum calculations. This level of accuracy signifies that the model is robust and can effectively be used for predicting $I_{ds}$-$V_{gs}$ curves. The predicted $I_{ds}$-$V_{gs}$ curves for all 100 transistors using PHVNN are shown in **Fig. S3**, verifying the prediction accuracy. The predicted curves show a high level of agreement with the quantum calculation results.

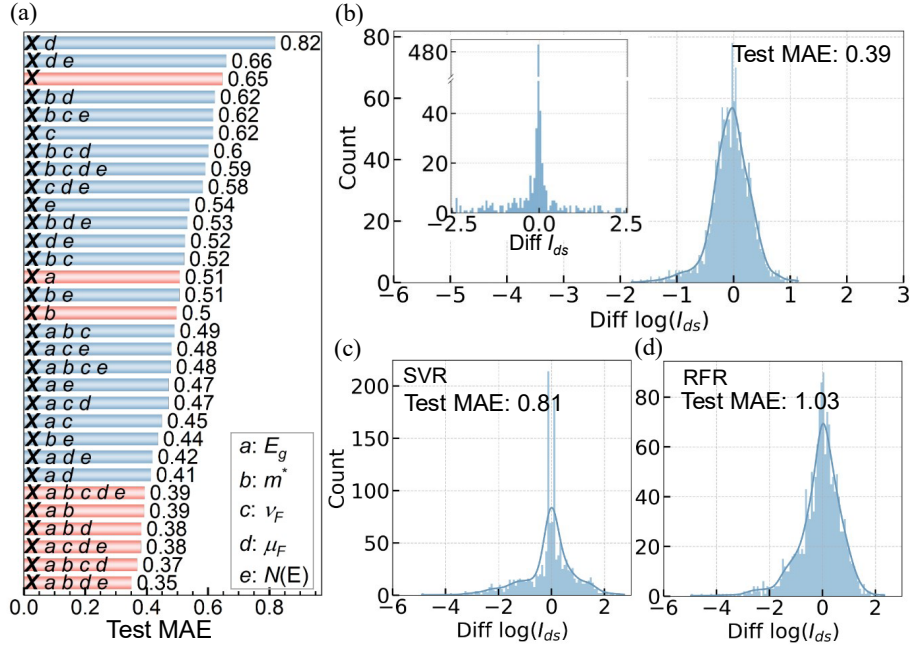

**Fig. 3.** (a) Ablation study in PHVNN for the contribution analysis of the five physical characters. The MAEs for test set are listed for each model in ablation study. *X* represents a group of the device size and transport setting representations (see details in Table 1). (b-d) Frequency distribution of the log($I_{ds}$) residuals between the predictions and the quantum calculation values for the PHVANN (b), SVR (c), and RFR (d). The residual distribution of the predicted $I_{ds}$ for the PHVANN is also provided (inset).

We then applied several classical ML models for the $I_{ds}$-$V_{gs}$ curve prediction as comparation. It is found that both random forest regression (RFR) and support vector regression (SVR) exhibit overfitting (Table S1) and much higher MAE than PHVNN (1.03 and 0.81 for RFR and SVR, respectively). Although kernel ridge regression avoids overfitting, it shows the worst MAE of 2.16. The log($I_{ds}$) residual distribution for RFR and SVR models (**Fig. 3 (c, d)**) also displays a large error range. The limitations and higher errors observed in classical ML models, like RFR and SVR, highlight the superiority of our PHVNN model in accurately predicting $I_{ds}$-$V_{gs}$ curves.

The symmetric *n*- and *p*-type FET pairs with the same figures of merit (FOMs) are required to achieve efficient logic operation. To identify symmetric *p*-type NW-FETs, we predicted the $I_{ds}$-$V_{gs}$ curves of *p*-type FETs with different electrode doping concentration, using InP GAA NW-FETs as an example. These predictions were made outside the training range, with all the InP NW-FETs configured with $L_g$ = 5 nm and $L_{UL}$ = 0 nm. It was predicted that the *p*-type InP FETs with $N_d$ of 2.2 × $10^{20}$ and 7.2 × $10^{19}$ cm$^{-3}$ can provide symmetry performance as the *n*-type InP FETs with $N_d$ of 3.6 × $10^{20}$ and 7.2 × $10^{19}$ cm$^{-3}$, respectively (**Fig. 4 (a, b)**). To validate the prediction capability of PHVNN, we performed NEGF calculations on the $I_{ds}$-$V_{gs}$ curves of the two *p*-types InP GAA FETs. **Figures 4a** and **4b** show a good agreement between the predicted $I_{ds}$-$V_{gs}$ curves and the values obtained from quantum calculations, further demonstrating the reliability and prediction accuracy of PHVNN. Moreover, we also validate our model using a few *n*- and *p*-type Se NW-FETs with $N_d$ of 1.0 × $10^7$ cm$^{-1}$, which were not used in model training process. **Figures 4a** and **4b** (orange curves) show that the *n*-type Se FETs with $L_g$ = 1 nm and $L_{UL}$ = 4 nm, and the *p*-type with $L_g$ = 5 nm and $L_{UL}$ = 0 nm, exhibits approximately symmetrical performance, demonstrating the potential of

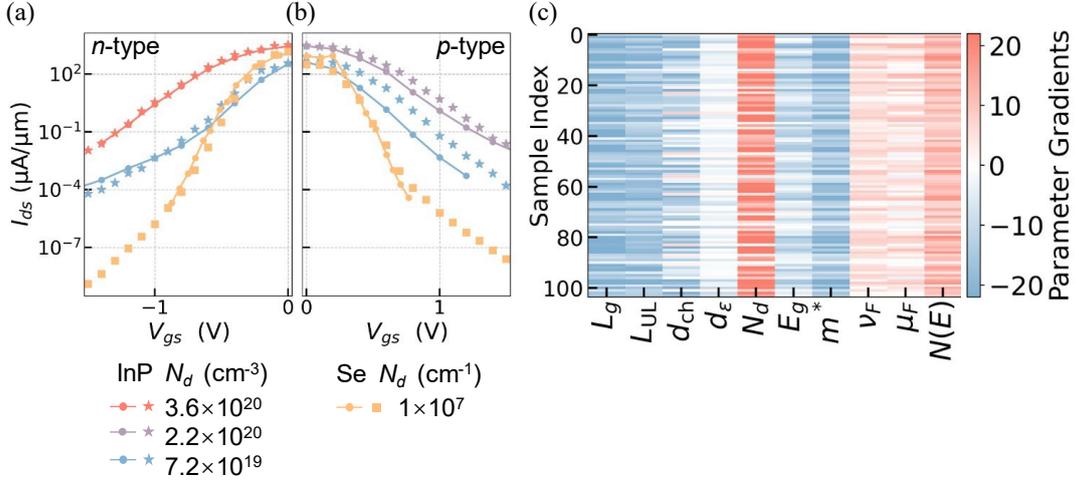

**Fig. 4.** (a, b) Predicted $I_{ds}$-$V_{gs}$ curves of the *n*- and *p*-type GAA WZ InP and Se NW-FETs. The cycle line symbols are the quantum calculation values, and the star or square are the PHVANN prediction results. All the InP NW-FETs and the *p*-type Se NW-FETs have the $L_g$ of 5 nm without underlap regions, but the *n*-type Se NW-FET has the $L_g$ of 1 nm and $L_{UL}$ of 4 nm. (c) Automatic differentiation for the output layer corresponding to the input features.

## 3. Physical Analysis

The ablation study highlights the importance of the five physical parameters (**Fig. 3(a)**), showing that these parameters contribute differently to the $I_{ds}$-$V_{gs}$ curve prediction. Notably, the model incorporating four parameters ($E_g$, $m^*$, $\mu_F$, and $N(E)$) achieved the lowest test MAE of 0.35, not using all 5 parameters. It is because the carrier Fermi velocity and effective mass of channel material are directly derived from the same band structure. Thus, considering both might introduce redundancy and increase the model's complexity. Interestingly, the model can achieve a test MAE of 0.39 using only two parameters of bandgap and effective mass, underscoring the significance of these two parameters in predicting the $I_{ds}$-$V_{gs}$ curve. The model using only $E_g$ (0.51) or $m^*$ (0.50) also has a low MAE, further emphasizing their critical roles in the prediction of $I_{ds}$-$V_{gs}$ curves.

Automatic differentiation (AD) is a core technology in physics-informed neural networks (PINNs)[37], which offers an effective method for numerically analyzing the derivatives of features embedded within a neural network. Since PHVNN exclusively utilizes dense NNs, the entire framework is infinitely differentiable. This property allows us to investigate the differentiation of the output with respect to the input parameters (**Fig. 4(c)**) to analyze their contributions to $I_{ds}$. As expected, gate length and UL negatively influence $I_{ds}$, while the doping concentration positively influences $I_{ds}$.

For nanometer FETs, the primary contributors to $I_{ds}$ are the addition of tunneling current ($I_{\text{tunnel}} \propto e^{-w\sqrt{\Phi m^*}}$) and the thermionic current ($I_{\text{therm}} \propto e^{-\Phi}$), where $\Phi$ is the carrier barrier height, and $w$ is the barrier width[2]. Our AD analysis confirms that $m^*$ negatively contributes to $I_{ds}$. The barrier width can be approximated by the effective screening length of the channel ($w \sim \lambda_{ch}$), which is influenced by the electrical permittivity ($\epsilon_{\text{ch}}$, $\epsilon_{\text{ox}}$) and the thickness ($d_{\text{ox}}$, $d_{\text{ch}}$) of the channel and dielectric oxide. According to the EOT expression (Method), $d_{\text{ox}}/\epsilon_{\text{ox}}$ can be replaced by the thickness of EOT, learning to $\lambda_{ch} \propto \sqrt{(\epsilon_{ch}/\epsilon_{SiO_2})\, d_\epsilon d_{ch}}$. It can be clearly seen that both $d_{ch}$ and $d_\epsilon$ negatively impact $I_{ds}$.

According to the definition of carrier Fermi velocity (Method), the barrier height is approximately equal to Fermi velocity ($-\Phi \approx \mu_F$), and thus, the $\mu_F$ is positively correlated with $I_{ds}$. Both $E_g$ and $N_d$ also relate to $\Phi$, which severely limits transistor conductance in the 'ON' state, reducing the current delivery capability—a key determinant of device performance. A larger bandgap typically results in a higher $\Phi$, leading to a smaller $I_{ds}$, while higher doping levels ($N_d$) push the Fermi level deeper into the conduction or valence band, increasing $\mu_F$ and thus $N_d$. Finally, carrier contribution states representing the density of states or transport channels, significantly contributes to carrier transport and consequently to $I_{ds}$. These results from physical analysis further demonstrate that our PHVNN model has effectively learned the underlying physical principles governing the transistor behavior. The alignment of the model's predictions with established physical knowledge highlights its interpretability and significantly enhances its credibility and reliability in predicting the device performance.

**Conclusion**

In conclusion, we have developed a physics-integrated neural network model for effectively predicting the quantum transport performance of sub-5 nm GAA NW-FETs, achieving a test MAE of 0.39 for log($I_{ds}$) hyper vectors, even at small-data limit. This high level of prediction accuracy is largely attributable to the rational selection of input parameters with prior physical knowledge, which are validated through an ablation study. Through the first-principles calculations and physical analysis, we demonstrate that our model's ability to incorporate and leverage these fundamental physical principles not only enhances its performance, but also significantly improves its interpretability and reliability. The PHVNN can visualize the importance of integrating prior-physical knowledge into neural networks. It achieves robust and reliable predictions in advanced semiconductor device research. The idea of PHVNN can also be extended in the future to predict additional performance metrics of ultra-short channel FETs, such as capacitance, delay time, and power dissipation.

**Methods**

All of the doped and undoped material physical parameters are calculated with the first-principles DFT method carried out in the Vienna *ab initio* simulation package (VASP) with the help of VASP high throughput calculation (HTC) framework (https://github.com/bitsoal/VASP_HTC_framework). The plane-wave cutoff energy is set to 650 eV, and a fine *k*-mesh density of 0.02 Å$^{-1}$ under the Monkhorst-Pack method is sampled in the Brillouin zone. The force tolerance on each atom is less than 0.01 eV/Å, and the converged energy is less than $10^{-6}$ eV. All of the 1D NW calculated with a supercell larger than 40 Å in the *x* and *y* axes, making the vacuum buffer space of at least 24 Å to avoid spurious interactions, and the *z* axis is set as the periodic direction.

The EOT is defined as $d_\epsilon = (\epsilon_{SiO_2}/\epsilon_{ox})d_{ox}$, where $\epsilon_{SiO_2}$ is the permittivity of silicon dioxide, and $\epsilon_{ox}$ and $d_{ox}$ are the permittivity and thickness of dielectric oxide, respectively. Depending on the doping type, the transport carriers can be electrons or holes, and the corresponding physical parameters will differ. For *n*-type doped electrodes, the $m^*$ and $v_F$ associated with the conduction band minimum (CBM) are used, as electrons predominantly contribute to the transport in the FETs.

$m^*$ is calculated from the band structure using the relation $\frac{1}{m^*} = \frac{1}{\hbar^2}\left(\frac{d^2E}{dk^2}\right)$. $v_F$ is approximated by the group velocity and is given by $v_F = \frac{1}{\hbar}\frac{dE}{dk}$. The Fermi level of electrodes will move up, and the carrier Fermi velocity is calculated relative to the CBM ($\mu_F = E_F - E_{\text{CBM}}$). Due to the applied bias voltage, the Fermi level shifts further upward, causing more carriers to participate in the transport process. $N(E)$ is integrated from the CBM to $E_F$ ($N(E) = \int_{\text{CBM}}^{\mu_F} DOS(E)dE$) including the effect of half the bias voltage. For *p*-type doping, the corresponding values would be $m^*$ and $v_F$ of holes, along with $N(E)$ and $\mu_F$ relative to the VBM.

**Acknowledgement:**

We thank Dr. Shiqi Liu and Dr. Qiuhui Li for fruitful discussions on the NWFETs database, and thanks Dr. Chaofei Liu for insightful suggestions throughout the writing of this manuscript. We acknowledge financial support from Singapore MOE Tier 2 (No. A-8001872-00-00), and Open Fund of Hubei Key Laboratory of Electronic Manufacturing and Packaging Integration (Wuhan University) (No. EMPI2024009). The calculation work was performed at National University of Singapore HPC and the High-performance Computing Platform of Peking University.